\documentclass[onecollarge,natbib]{svjour2}
\bibpunct{[}{]}{;}{n}{}{,} % to get "[numbered]" references from natbib
\smartqed  % flush right qed marks, e.g. at end of proof
\usepackage{graphicx}
\journalname{Few Body Systems}
\begin{document}

\title{Nonperturbative calculations in truncated Fock space in LFD%\thanks{Grants or other notes
}


\author{V.A.~Karmanov        
% \and        Second Author %etc.
}


\institute{V.A.~Karmanov \at
              Lebedev Physical Institute, Leninsky Prospect 53, 119991 Moscow, Russia \\
%              Tel.: +123-45-678910\\
 %             Fax: +123-45-678910\\
              \email{karmanov@sci.lebedev.ru}           %  \\
 %             second address
}

\date{Received: date / Accepted: date}

\maketitle

\begin{abstract}
A non-perturbative approach based on the Fock decomposition of the state vector and its truncation is discussed.
In order the non-perturbative renormalization procedure after truncation could eliminate infinities, it should be the 
sector dependent. We clarify the meaning of this procedure in a toy model.
Then we demonstrate stability, relative to the increasing cutoff, of the anomalous 
magnetic moment found using the sector dependent renormalization scheme in Yukawa model.

% \and More}
\end{abstract}

\section{Introduction}
\label{intro}
%Your text comes here. Separate text sections with
Any field-theoretical Hamiltonian does not conserve the number of particles. 
Therefore, in the basis, corresponding to fixed number of particles, it is a non-diagonal matrix. 
Its eigenvector -- the state vector of a physical system -- is an infinite  superposition (Fock decomposition) of the states with different numbers of particles:
\begin{equation}\label{eq1}
\left\vert p \right> = \sum_{n=1}^{\infty} \int
\psi_n(k_1,\ldots,k_n,p)\left\vert n \right>D_k.
\end{equation}
$\psi_n$ is the $n$-body wave function (Fock component) and $D_k$ is an integration measure.

In many cases, though not always, we can expect that a finite number of degrees of freedom dominates, that is, the decomposition 
 (\ref{eq1}) converges enough quickly. In some examples the convergence is even better than one can expect naively. 
 In these cases  we can  make truncation, that is replace the infinite sum in (\ref{eq1}) by the finite one. 
Then, substituting truncated state vector in the eigenvector equation
$$
H\left\vert p \right>=M\left\vert p \right>,
$$
we obtain a finite system of integral equations for the Fock components $\psi_n$ which can be solved numerically. 
We do not require  the smallness of the coupling constant. The approximate (truncated) solution is non-perturbative. This is the basis of non-perturbative approach which we developed, together with  J.-F.~Mathiot and A.V.~Smirnov, 
in a series of our papers \cite{gauge,kms07,kms08,kms10,kms12} (see for review \cite{mstk}).

The main difficulty in this way is to ensure cancellation of infinities after renormalization. 
In perturbative approach, for a renormalizable field theory, in any fixed order of coupling constant, this cancellation, after renormalization,  is obtained 
as a by-product. However, 
it is important to take into account full set of graphs in a given order. Omitting some of these graphs destroys the cancellation and the infinities survive after renormalization. Namely that happens after truncation: though the truncated solution can be decomposed in infinite series in terms of the coupling constant, in any given order it does not contain full set of perturbative graphs. Therefore the standard renormalization scheme does not eliminate infinities. To restore cancellation of infinities, there was proposed \cite{phw1990} the sector-dependent 
scheme.
This scheme,  in which the values of the counter terms are precised from sector to sector according to unambiguously formulated rules, was developed in detail in \cite{kms08,kms10,kms12}. 

Following these rules, the problem, at first, should be solved and the counter terms are found in 
the two-body truncation. 
In the highest (two-body) sector the counter terms do not appear. 
Their presence would mean implicit incorporation of extra intermediate states. 
The counter terms correspond to sum of the graphs containing the intermediate particles (loop graphs, for example).  These states, together with the two-body ones, constitute the three-body sector and therefore they
exceed the two-body truncation. The one-body sector contains the two-body counter terms. 
They are found from the renormalization conditions imposed on the two-body solution. 

Then the problem is solved again, in the three-body truncation, which retains now the sum of one-, two- and three-body sectors.
In the highest (three-body) sector the counter terms do not appear.
The two-body sector contains already known counter terms found previously in the two-body truncation.
The one-body sector contains the three-body counter terms which appear there for the first time. They are found from the renormalization conditions imposed on the three-body solution. 

Then this procedure is repeated for the next truncation. We repeat it for clarity for the four-body truncation. Namely, 
the four-body truncation retains the sum of one-, two-, three- and four-body sectors.
In the four-body sector the counter terms do not appear (to avoid the exceed up to the five-body sector). The three-body sector contains the counter terms found previously in the two-body truncation. The two-body sector contains the counter terms found previously in the three-body truncation. The one-body sector contains the four-body counter terms which appear there for the first time. They are found from the renormalization conditions imposed on the four-body solution.  Etc.

When the number $N$ of incorporated Fock sectors increases,  it is naturally expected that the solutions, found in this way,  -- the state vector and the counter terms, -- converge to a limiting exact values.  However, this has never been checked. There are two reasons for that. ({\it i}) It is not easy to solve the equations for the Fock components for enough large $N$. For the present, it was solved for $N\leq 3$. ({\it ii}) There is no any field-theoretical model  in which the exact state vector and the counter terms are known. Therefore one cannot compare a truncated solution with the exact one.

We will give here a simple-minded example of the solvable "0-dimentional" field theory in which  ({\it i}) the equations for the Fock components for enough large $N$ can be solved numerically; ({\it ii})  the comparison of truncated solution with the exact one can be done. Our aim is two-fold. ({\it i}) To demonstrate in a simple example the sector dependent renormalization procedure.  ({\it ii})  To check the convergence of the truncated solutions for increasing $N$ to the exact  solution.
\vspace{-0.3cm}
%%%%%%%%%%%%%%%%%%%%%%
\section{Zero-dimensional model}
\vspace{-1cm}
\begin{figure}[h!]
%\begin{center}
\includegraphics[width=10cm]{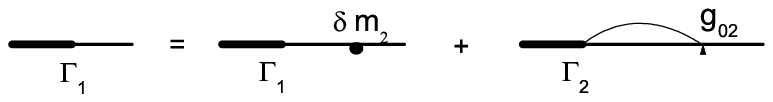}
\vspace{-1cm}\\
\includegraphics[width=10cm]{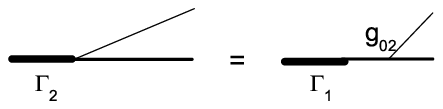}
%\end{center}
\vspace{-0.5cm}
\caption{System of equations in the two-body truncation}
\label{fig1}
\end{figure}

System of equations in the two-body truncation, relating one- and two-body components, is graphically shown in fig. \ref{fig1}.
$\Gamma_{1,2}$ are the vertex functions related to the wave functions  $\psi_{1,2}$ as
$$
\psi_1= \frac{\Gamma_1}{m^2-M^2},\quad \psi_2=\frac{\Gamma_2}{s-M^2}
$$
Deriving these equations, we first suppose that the bare mass $m$ (internal particle) and the external mass $M$ are not equal to each other. 
Then we impose on the mass counter term $\delta m$ the renormalization condition due to which $M\to m$.
The first term in the r.h.-side of the equation shown in the first line of fig. \ref{fig1}, contains the one-body propagator
corresponding to the line connecting $\Gamma_1$ and the vertex $\delta m_2$. The same propagator is also put in correspondence to the line connecting $\Gamma_1$ and $g_{02}$ in r.h.-side of the equation shown in the second line of fig. \ref{fig1}.
It just reads $1/(m^2-M^2)$ and it is absorbed into $\Gamma_1$, giving $\psi_1$. Whereas l.h.-side of this equation does not contain propagator. In terms of $\psi_1$ it obtains the form  $(m^2-M^2)\psi_1$ and disappears when we take $M=m$ keeping $\psi_1$ finite. 
Therefore the system of equations shown in fig. \ref{fig1} reads:
\begin{eqnarray}
0&=& \delta m_2 \psi_1 +V_{12}\psi_2
\nonumber\\
\psi_2&=& V_{21} \psi_1+\delta m_1\psi_2
\label{eq1a}
\end{eqnarray}

In the second term in r.h.-side of the first equation we replaced the integral term $\int \Gamma_2\ldots$ corresponding to the loop  by the product
$V_{12}\psi_2$. 
We will make similar replacements also in the equations for higher truncations. This is the reason why we call
this toy model the "zero-dimensional" model.  Also, for simplicity of notations, for $n\ge 2$,  we identify  here and below $\Gamma_n$ with $\psi_n$.
 
For clarity of construction, we keep in r.h.-side of the second equation in (\ref{eq1a}) the term $\delta m_1\psi_2$ which should be found from the previous one-body truncation. 
However, the one-body truncation is trivial -- it does not contain interaction. The latter changes the number of particles and therefore relates the one-body and two-body sectors. Therefore $\delta m_1=0$ and in the sector-dependent renormalization scheme this term does not contribute. That's why it is not shown in fig. \ref{fig1}. 

In the realistic case, the goal of the sector-dependent scheme is elimination of infinities, in spite of truncation. However, the infinities are absent in the 0D model.
We develop this (non-divergent) toy model in order to illustrate, in a simple example, the sector dependent renormalization procedure (even without eliminating infinities).
We will also check, when the truncation $N$ increases, whether the solution found in this scheme tends at all to the exact solution, and if it tends to it -- 
how quickly.  

Exact solution is the solution of the original field-theoretical equations in which the counter terms, in contrast to the sector dependent scheme, are the same in any sector and the dimension of the matrix acting on the (infinite) Fock column is infinite. To find the solution for infinite matrix, we still start with a finite matrix of the dimension $N\times N$, keeping the counter terms as they are (i.e. the same in all the sectors) and then take the limit $N\to\infty$. We start with the case $N=2$. The corresponding system of equations is obtained from (\ref{eq1a}) by setting $\delta m_1=\delta m_2=\delta m$. In the matrix form it reads:
\begin{equation}
\left(
\begin{array}{cc}
\delta m & V_{12}
\\
V_{21} & \delta m -1
\end{array}
\right)
\left(
\begin{array}{c}
\psi_1
\\
\psi_2
\end{array}
\right)
=0
\label{eq2}
\end{equation}

In the case $N=3$ the system of equations (\ref{eq1a}) is generalized as:
\begin{eqnarray}
0&=& \delta m \psi_1 +V_{12}\psi_2
\nonumber\\
\psi_2&=& V_{21} \psi_1+\delta m\psi_2 +V_{23}\psi_3
\nonumber\\
\psi_3&=&\;\phantom{V_{21} \psi_1+ } V_{32}\psi_2+ \delta m\psi_3
\label{eq1b}
\end{eqnarray}

The generalization of this system of equations to the case of matrix of arbitrary dimension has the form:
\begin{equation}
\sum_i M_{ij}\psi_j=0, \quad
\mbox{where}
\quad
M_{ij}=V_{ij}+\delta_{ij}(\delta\,m-1+\delta_{i1}\delta_{j1})
\label{eq3}
\end{equation}
and interaction $V_{ij}$ should satisfy  two following properties: ({\it i}) it connects only the neighbor  components: 1-body $\leftrightarrow$ 2-body,  2-body $\leftrightarrow$ 3-body, etc., like in the case of the interaction $g\phi^3$ or  the Yukawa model $g\psi\bar{\psi}\phi$;
({\it ii}) it becomes weaker for higher components (since creation of large number of particles requires more energy). The latter property should be also automatically provided by a field-theoretical Hamiltonian; in the toy model we mimic it by constructing $V_{ij}$ which decreases when $i,j$ increase.
The interaction $V_{ij}$, satisfying these properties, can be chosen, for example, as:
 \begin{equation}\label{Vij1}
V_{ij}= \frac{g}{(i+j)}\Delta_{ij},
\quad 
\mbox{where
$
\Delta_{ij}=\left\{\begin{array}{l}
0, \quad \mbox{if $i=j$}
\\
1,\quad \mbox{if $|i-j|= 1$}
\\
0,\quad \mbox{if $|i-j|> 1$}
\end{array}
\right.
$}
\end{equation}

We will find at first the exact solution. We take strong coupling constant $g=2$.
The equation (\ref{eq2}) (the $N=2$ case) obtains the form:
\begin{equation}
M\psi=\left(
\begin{array}{cc}
\delta m & \frac{2}{3}
\\
\frac{2}{3} & \delta m -1
\end{array}
\right)
\left(
\begin{array}{c}
\psi_1
\\
\psi_2
\end{array}
\right)
=0.
\label{eq4}
\end{equation}
Solving the quadratic equation $det(M)=0$ relative to $\delta m$, we find two solutions: 
$\delta m=-0.33$, $\delta m=1.33$. 

In the case $N=3$, solving corresponding cubic equation $det(M)=0$ with $M$ defined in (\ref{eq1b}) (or, equivalently, in (\ref{eq3}), (\ref{Vij1})), 
we find three solutions: 
$\delta m=-0.358$, $\delta m=0.792$, $\delta=1.566$.  We will consider the first (negative) solution
as the physical one. Increasing $N$ up to $N=10$ (and solving numerically the equations for $\delta m$ up to the 10th degree), we find the values of $\delta m$ shown in the table \ref{tab1}. The digits which remain
stable when $N$ increases are underlined. For example, the underlined digits in $\delta m = -\underline{0.35}82$ for $N=3$ (i.e., $-\underline{0.35}$) are reproduced in the value $\delta m=-\underline{0.359}48$ for $N=4$. The underlined digits in $\delta m$ for $N=4$ (i.e., $-\underline{0.359}$) are reproduced for $N=5$.  The convergence of the $\delta m$ value when $N$ increases is very fast. For $N=9$ we get 14 digits which are reproduced in the next truncation $N=10$. The solution with the precision 10 digits (for $N=8$) or 14 digits (for $N=9$) we call the "exact" solution. Strictly speaking, it is not exact, but the precision 
in 10-14 digits is quite enough.

\begin{table}[h]%[btph]
\caption{The value of $\delta\, m$ found by solving eq. (\ref{eq3}), with the kernel (\ref{Vij1}), with the matrix $M$ truncated up to the $N\times N$ dimension for $N=2,3,\ldots,10$.}
\centering
\label{tab1}
\begin{tabular}{rl}
\hline\noalign{\smallskip}
$N$            & $\phantom{-}\delta\, m$\\[3pt]
\tableheadseprule\noalign{\smallskip}
1  & $\phantom{-}0$\\
2  & $-\underline{0.3}33$\\
3  & $-\underline{0.35}82$\\
4  & $-\underline{0.359}48$\\
5  & $-\underline{0.35951}54$\\
6  & $-\underline{0.35951607}02$\\
7  & $-\underline{0.359516078}79$\\
8  & $-\underline{0.3595160788}796$\\
9  & $-\underline{0.35951607888029}41$\\
10 & $-0.3595160788802980$\\
\noalign{\smallskip}\hline
\end{tabular}
\end{table}

For the case $N=8$ we substitute the eigenvalue $\delta m =-0.3595160788796$ into the matrix (\ref{eq3}) and find 
the corresponding eigenvector:
\begin{equation}\label{psi8a}
\psi=\left(\begin{array}{l}
0.870\\
0.469\\
0.145\\
0.312\cdot 10^{-1}\\
0.520\cdot 10^{-2}\\
0.705\cdot 10^{-3}\\
0.806\cdot 10^{-4}\\
0.790\cdot 10^{-5}
\end{array}
\right)
\end{equation}
We normalized it to 1: $\langle\psi\vert\psi\rangle= \psi_1^2+\psi_2^2+\ldots +\psi_8^2=1$. The first three components dominate: they give $99.8\%$ of full normalization. This justifies the truncation.

Now let us use the sector-dependent scheme. 
This means that in the system of equations (\ref{eq1a})  we put $\delta m_1=0$ and find $\delta m_2$. 
Instead of (\ref{eq4}) we get: 
\begin{equation}
M\psi=\left(
\begin{array}{cc}
\delta m_2 & \frac{2}{3}
\\
\frac{2}{3} & -1
\end{array}
\right)
\left(
\begin{array}{c}
\psi_1
\\
\psi_2
\end{array}
\right)
=0.
\label{eq4a}
\end{equation}
Solving equation $det(M)=0$  (which is now linear relative to $\delta \, m_2$)
we find $\delta \, m_2=-4/9=-0.444$,
 in comparison to the $N=2$ value $\delta m= -0.333$ from the table \ref{tab1}.

In the case $N=3$ the system of equations (\ref{eq1b}) is replaced by:
\begin{eqnarray}
0&=& \delta m_3 \psi_1 +V_{12}\psi_2
\nonumber\\
\psi_2&=& V_{21} \psi_1+\delta m_2\psi_2 +V_{23}\psi_3
\nonumber\\
\psi_3&=&\;\phantom{V_{21} \psi_1+ } V_{32}\psi_2+ \delta m_1\psi_3
\label{eq1c}
\end{eqnarray}
where $\delta \, m_1=0$, $\delta \, m_2=-4/9$ and $\delta m_3$ is the mass counter term for the $N=3$ sector to be found.
The matrix equation obtains now the form:
\begin{equation}
M\psi=
\left(\begin{array}{ccc}
\delta\, m_3 & V_{12} & 0
\\
V_{21} & \delta\, m_2 -1 & V_{23}
\\
0 & V_{32} & \delta m_1-1
\end{array}
\right)
\left(\begin{array}{c}
\psi_1
\\
\psi_2
\\
\psi_3
\end{array}
\right)=
\left(
\begin{array}{ccc}
\delta m_3 & \phantom{-}2/3 & \phantom{-}0 
\\
2/3  &  -13/9 & \phantom{-}2/5 \\
0 &  \phantom{-}2/5 & -1
 \end{array}
 \right)
\left( \begin{array}{c}
\psi_1
\\
\psi_2
\\
\psi_3
\end{array}
\right)
=0
\label{eq6}
\end{equation}
From the linear equation $det(M)=0$ we find
$\delta \, m_3=-100/289=-0.346021$, in comparison to the $N=3$ value $\delta m= -0.3582$ from the table \ref{tab1}.

From eq. (\ref{eq6}) one can already guess that, 
in general, in the sector-dependent scheme the matrix
$M$ in (\ref{eq3}) is replaced by the following $N\times N$ matrix:
\begin{equation}
M_{ij}=V_{ij}+\delta_{ij}(\delta\,m_{N+1-i}-1+\delta_{i1}\delta_{j1})
\label{eq5a}
\end{equation}
with $\delta\,m_{N+1-i}$ found successively from sector to sector (and $\delta m_1=0$). 

Solving the equation $det(M)=0$ (which is still linear relative to $\delta m_N$) with the values $\delta m_1=0$, and $\delta m_2,\ldots, \delta m_{N-1}$ 
found previously, 
we  find successive sector-dependent values of $\delta\, m_N$ shown in the table \ref{tab4}. Like in the table \ref{tab1}, the digits which remain stable 
when $N$ increases are underlined.
The 4th iteration $\delta \, m_4=-0.3617$ gives the precision $0.5\%$. The 8th iteration $\delta \, m_4=-0.359517$  gives the precision of the order of $10^{-6}$.
\begin{table}[t]%[btph]
\caption{The value of $\delta\, m_N$ found by solving the successive set of equations (\ref{eq4a}),   (\ref{eq1b}) and, in general, $det(M)=0$ with $M$ defined in (\ref{eq5a}) and with the kerne $V_{ij}$ defined in (\ref{Vij1}), in comparison to the exact value from the line $N=8$ in the table \ref{tab1}.}
\centering
\label{tab4}
\begin{tabular}{rl}
\hline\noalign{\smallskip}
$N$            & $\phantom{-}\delta\, m_N$\\[3pt]
\tableheadseprule\noalign{\smallskip}
1  & $\phantom{-}0$\\
2  & $-0.444$\\
3  & $-\underline{0.3}46$\\
4  & $-\underline{0.3}617$\\
5  & $-\underline{0.359}15$\\
6  & $-\underline{0.3595}74$\\
7  & $-\underline{0.3595}06$\\
8  & $-\underline{0.35951}7$\\
exact
   & $-0.3595161$\\
\noalign{\smallskip}\hline
\end{tabular}
\end{table}

We have also considered the case of very strong coupling constant $g=8$ in the kernel (\ref{Vij1}). In this case $\delta m_N$ still converges to the exact value 
though more slowly. In the case of quickly decreasing kernel $V_{ij}$ relative to increase of $i,j$, namely 
$$
V_{ij}= \frac{2^4}{(i+j)^4}\Delta_{ij}
$$
the convergence is super fast. In the sector dependent scheme the value $N=8$ provides $\delta m_N$, coinciding with the exact one, with the  stability in 40 digits.

This simple example illustrates the iterative procedure which is used in the sector-dependent renormalization scheme. 
It shows that the mass counter $\delta m_N$, found by this procedure from a linear equation,  converges, when $N$ increases, rather quickly to the "exact" value calculated without any sector-dependent scheme. 

In this toy model, the renormalization of the coupling constant is absent.  In the realistic case, it appears. In addition,  another counter term $Z_{\omega}$ appears \cite{kms12}, which eliminates, on the mass shell,  in the $2\times 2$-matrix, representing the two-body vertex in the spinor basis,
the non-diagonal (light-front orientation dependent) elements. After that this vertex can be identified with the coupling constant.
However,  the sector dependent renormalization procedure remains the same: instead of one counter term $\delta m_N$, we should find now from the renormalization conditions, for given $N$, the three ones: still $\delta m_N$, the bare coupling constant $g_{0N}$ (i.e., express it in terms of the physical one $g$) and also $Z_{\omega,N}$.
Then we use these three values in calculations in the next $N+1$ sector.  

%%%%%%%%%%%%%%%%%%%%%%%
\section{Yukawa model}
As mentioned, the sector dependent renormalization procedure is aimed to cancellation of infinities, for any given truncation $N$. This example does not show this property, 
since the infinities are absent at all. Similar calculation in Yukawa model, containing divergences and renormalization of mass and the coupling constant,
was carried out in \cite{kms12} in the $N=3$ truncation  for  three values of the physical coupling
constant, $\alpha=g^2/4\pi=0.5$, $0.8$, and $1.0$. The Pauli-Villars  (PV) regularization with one PV fermion and one PV boson was used. If the infinities are indeed cancelled, the renormalized results should not depend on the values of the PV masses when the latter ones tend to infinity. 

\begin{figure}[ht!]
\begin{center}
\includegraphics[width=0.45\textwidth]{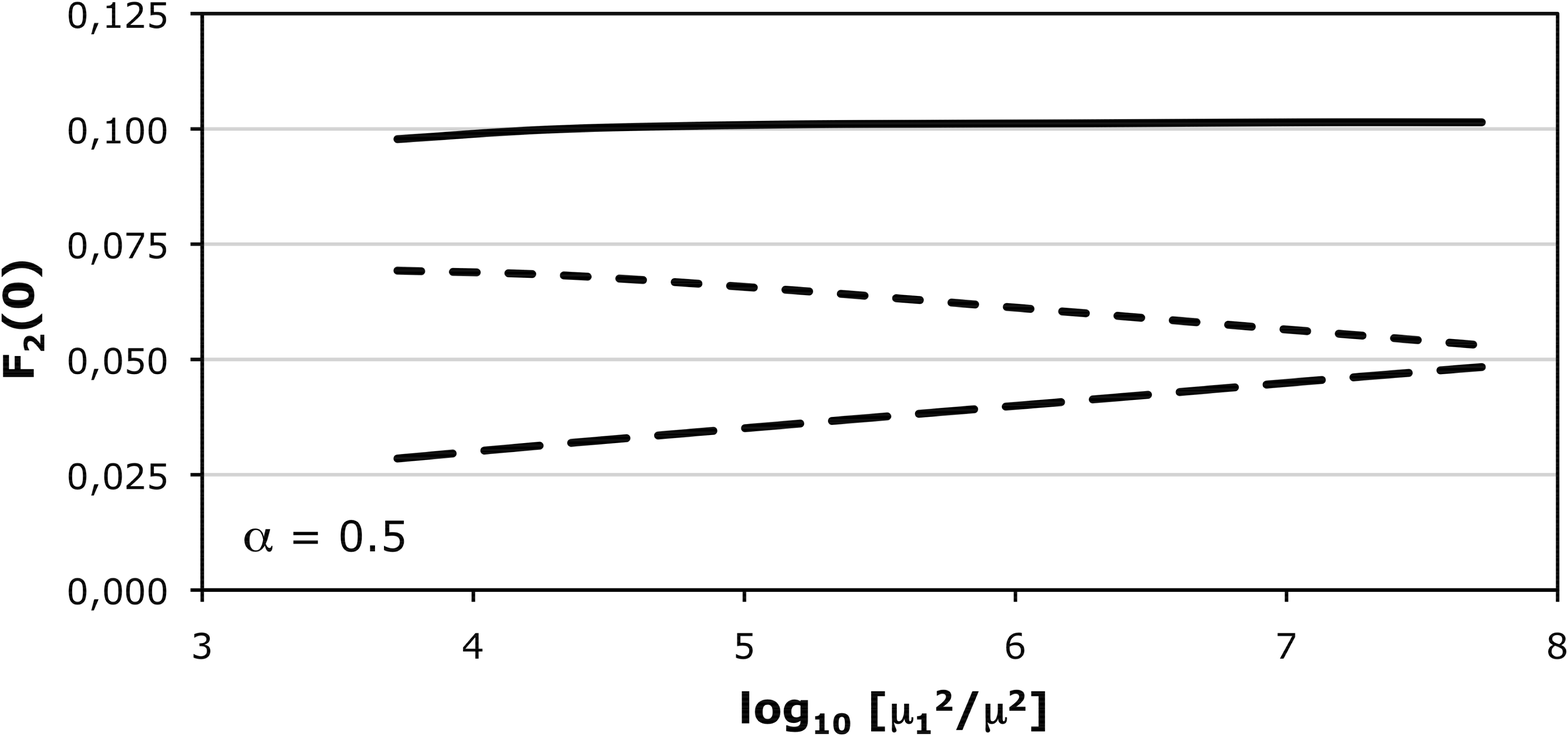}
\includegraphics[width=0.45\textwidth]{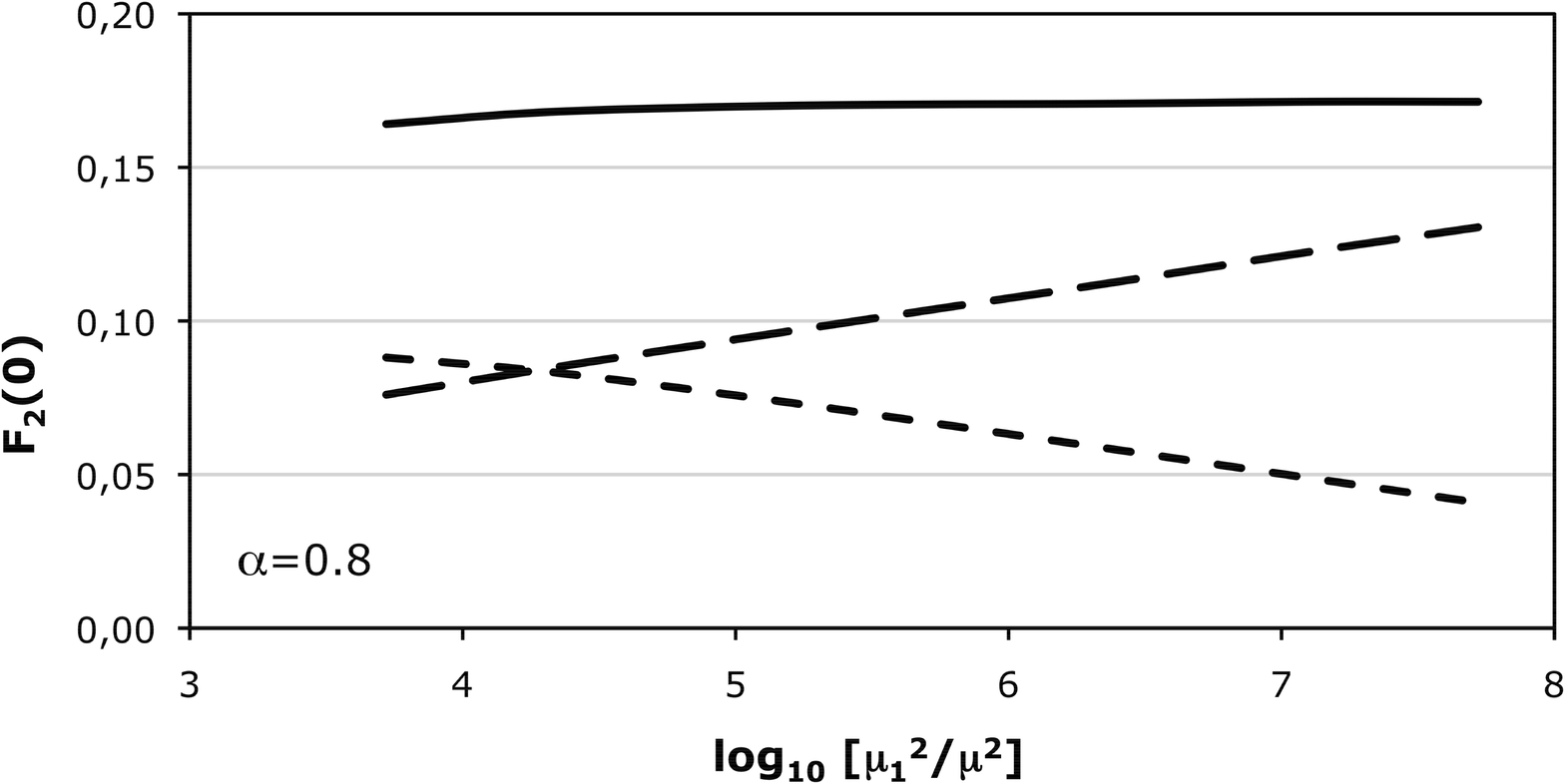}\\
\includegraphics[width=0.45\textwidth]{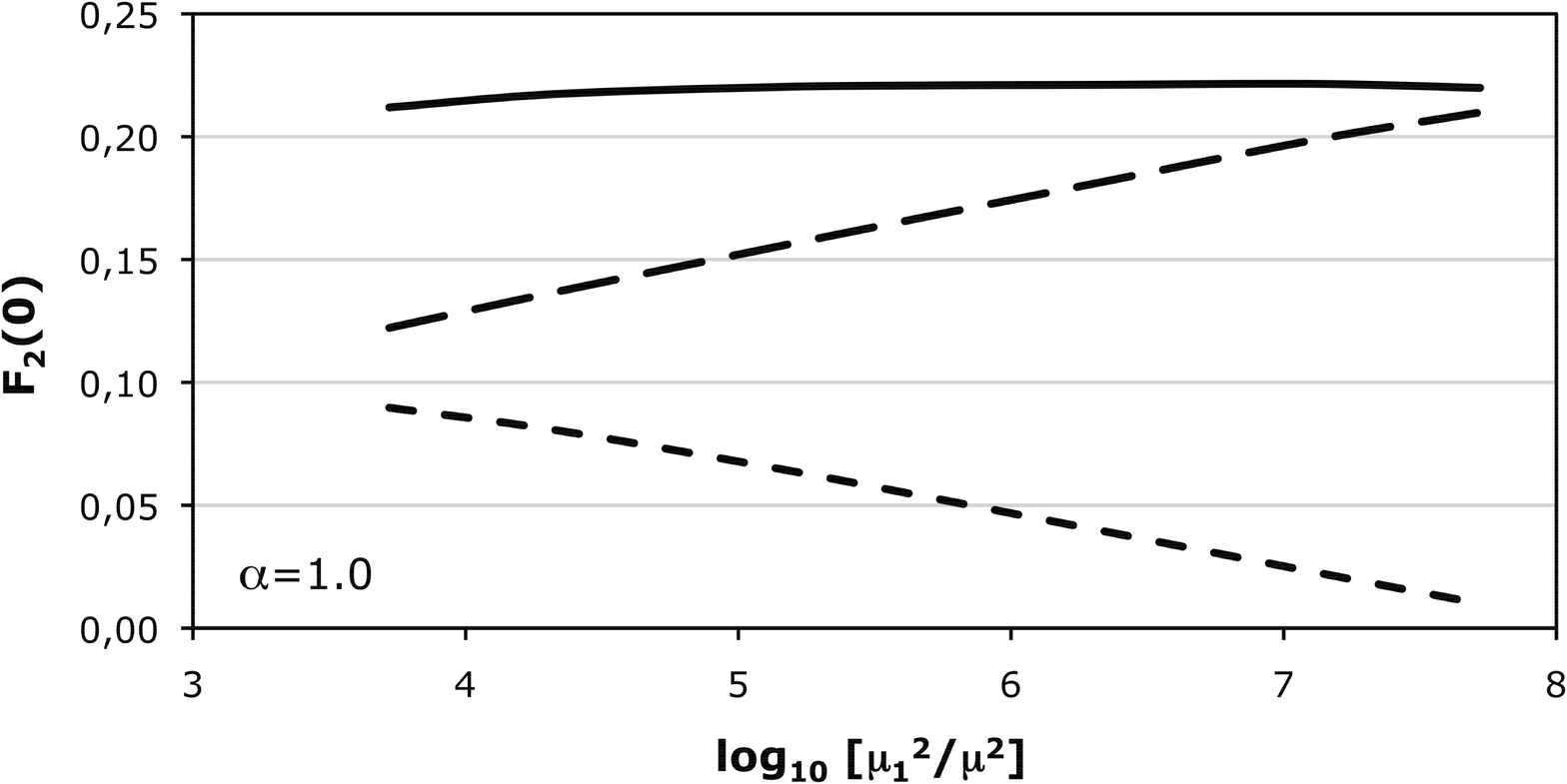}
\end{center}
\caption{The anomalous magnetic moment in the Yukawa model, calculated in \cite{kms12}, as a
function of the boson Pauli-Villars mass $\mu_1$, for three
different values of the coupling constant, $\alpha = 0.5$ (upper left
plot), $0.8$ (upper right plot) and $\alpha = 1.0$ (lower plot). The
dashed and long-dashed lines are, respectively, the two- and three-body
contributions, while the solid line is the total
result.}\label{amm}
\end{figure}

The anomalous magnetic moment, which is 
the value of the electromagnetic form factor $F_2(0)$,
is shown in  Fig.~\ref{amm} as a function of the PV boson mass $\mu_1$.
The limit $m_1\to\infty$ of the fermion PV mass was taken analytically.
One can see that each of the
two- and three-body sector contributions to the anomalous magnetic moment
depends on $\mu_1$,  while their sum is stable as
$\mu_1$ becomes large enough.

This stability indicates that in the sector-dependent renormalization scheme the infinities are cancelled, as expected. 
The check of stability in calculations in the $N=4$ truncation, as well as finding an indication on possible saturation of the results when $N$ increases from $N=3$ to $N=4$, first with the spineless particles and then in the Yukawa model, would be very interesting. 

\end{document}